\documentclass[pra,twocolumn]{revtex4}
\usepackage{bm,graphicx,amsmath}

\newcommand{\bra}[1]{\ensuremath{\langle#1|}}

\newcommand{\ket}[1]{\ensuremath{|#1\rangle}}

\newcommand{\braket}[2]{\ensuremath{\langle #1|#2\rangle}}

\begin{document}

\title{A deterministic cavity-QED source of polarization entangled photon pairs}
\author{R.~Garc\'{i}a-Maraver, K.~Eckert, R.~Corbal\'{a}n, and J.~Mompart}
\affiliation{Departament de F\'{i}sica, Universitat Aut\`{o}noma de Barcelona, E-08193 Bellaterra, Spain}
\date{\today}

\begin{abstract}

We present two cavity quantum electrodynamics proposals that, sharing the same basic elements, allow for the deterministic generation of entangled photons pairs by means of a three-level atom successively coupled to two single longitudinal mode high-Q optical resonators presenting polarization degeneracy. In the faster proposal, the three-level atom yields a polarization entangled photon pair via two truncated Rabi oscillations, whereas in the adiabatic proposal a counterintuitive Stimulated Raman Adiabatic Passage process is considered. Although slower than the former process, this second method is very efficient and robust under fluctuations of the experimental parameters and, particularly interesting, almost completely insensitive to atomic decay. 

\end{abstract}

\maketitle

The main issue in cryptography is the secure distribution of the encoding key between two partners. With this aim, quantum cryptography renders two classes of protocols \cite{BB84,SARG04,Ek91} based, respectively, on superposition and quantum measurement, or entanglement and quantum measurement. Entanglement based protocols were first considered by A.~Ekert \cite{Ek91} and present some potential advantages: (i) under passive state preparation, frustration of multiphoton splitting attacks since each photon pair is uncorrelated from the rest \cite{BLM00}; (ii) in the presence of dark counting, entangled states allow for the detection and removal of empty photon pulses by means of coincidence photodetection; (iii) for some entangled states lying in decoherence free subspaces, no information flows to non-relevant degrees of freedom \cite{MY98}; and, for quantum networks, (iv) null information leakage to the provider of the key. In spite of (i), it is very important to deal with single photon pairs since multiphoton pairs decrease the quantum correlations between the measurement results of the two parties and, accordingly, enhance the quantum bit error rate \cite{RBG01}.  

Quantum cryptography with entangled states has been achieved by means of parametric down converted photons generated in non-linear crystals \cite{QCE1,QCE2,QCE3}. However, in all these cases the photon number statistics (and their time distributions) follows, essentially, a poissonian distribution. Thus, in order to reduce the number of multiphoton pairs, the average photon number has to be much less than one which, in turn, strongly reduces the key exchange rate. Accordingly, one of the practical issues in entanglement based quantum cryptography presently attracting considerable attention is the development of light sources that emit deterministically single entangled photon pairs at a constant rate \cite{ZSS05,YYG05}. In addition, it is worth to notice that single pairs of entangled photons  and more involved non-classical photon states \cite{Mor05}
have a fundamental significance for testing quantum mechanics against local hidden variable theories and for practical applications in teleportation \cite{teleportation} and dense coding \cite{coding}.

Focusing on the optical regime, we discuss here a cavity-quantum electrodynamics (cQED) implementation \cite{Har,Wal,Kim,Oro,Kuh,Hei,bestatomic,Sauer} that, making use of a $V$-type three-level atom coupled successively to two high-Q cavities presenting polarization degeneracy, allows for the deterministic generation of polarization entangled photon pairs. The initial separable state of the system is chosen such that the relevant couplings can be reduced to those of a three-level interaction between the initial state and a bright state combination \cite{Ari} of the two excited atomic states and the modes of cavity 1, and between this bright state and the polarization entangled photon state. Two different scenarios are investigated for the entangled photon pair generation based, respectively, on two truncated Rabi Oscillations (ROs) and on a Stimulated Raman Adiabatic Passage (STIRAP) process \cite{STIRAP} between the three relevant states of the system. The feasibility of both implementations and some practical considerations will be discussed for realistic parameter values of state of the art experiments in optical cQED \cite{Kim,Oro,Kuh,Hei,bestatomic,Sauer}.  

The system under investigation is sketched in Fig.~1 and is composed of a single $V$-type three-level atom with two electric dipole transitions of frequencies $\omega_{ac}$ and $\omega_{bc}$, and two high-Q cavities both displaying polarization degeneracy and having identical longitudinal mode frequency $\omega_c$. $\Delta_{+} = \omega_c - \omega_{ac}$, and $\Delta_{-}= \omega_c - \omega_{bc}$ are the detunings. The transition $\ket{a,n_{i+}}\leftrightarrow \ket{c,n_{i+}+1}$ ($\ket{b,n_{i-}}\leftrightarrow \ket{c,n_{i-}+1}$) will be governed by the coupling $g_{i+}\sqrt{n_{i+}+1}$ ($g_{i-}\sqrt{n_{i-}+1}$) with $i=1,2$ denoting the cavity, $g_{i\pm}$ the vacuum Rabi frequency of the corresponding circular polarization, and $n_{i\pm}$ the number of $\sigma_\pm$ circularly polarized photons.   
We will consider here the completely symmetric case given by $\omega_{ac}=\omega_{bc}$, $\Delta_{+} =\Delta_{-} (\equiv \Delta )$, and $g_{i+}(t)=g_{i-}(t) (\equiv g_i (t)) $. This symmetry could be easily obtained by considering a $J=0$ $\leftrightarrow$ $J=1$ atomic transition with the quantization axis along the optical axis of the cavities. Eventually, we will relax some of the previous symmetry requirements in analyzing the influence of experimental imperfections such as the presence of a stray magnetic field. 

\begin{figure}[h]
	  \centering
	 	\includegraphics[width=\columnwidth,clip=true]{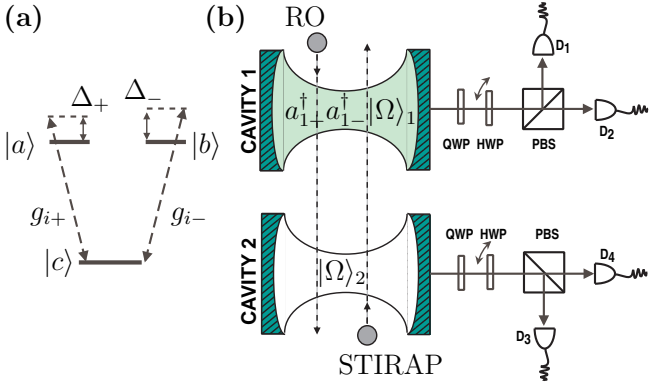}
\vskip -5mm
		\caption{(a) $V$-type three level atomic configuration under investigation. (b) Proposed setup for the deterministic generation of polarization entangled photon pairs. $\ket{\Omega}_i$ is the two mode vacuum state of cavity $i$. The two proposed configurations based, respectively, on two truncated ROs and STIRAP are simultaneously shown. Also shown are the basic optical elements needed for the Bell analysis of the entangled states (QWP: quarter wave plate; HWP: rotating half wave plate; PBS: polarization beam splitter; D$_i$: single photon detector).}
	\label{fig:f1ab}
\end{figure}

In the rotating-wave approximation, the coherent dynamics of the full system is described by the following Hamiltonian ($\hbar = 1$):
\begin{eqnarray}
H   &=&H_0+H_I,    \label{eq1} \\
H_0 &=& \sum_{i=1,2}  \omega_c \left( a^{\dag}_{i+} a_{i+} + a^{\dag}_{i-} a_{i-} \right)
+ \sum_{j=a,b} \omega_{jc}\ket{j}\bra{j}, \\
H_I &=& \sum_{i=1,2} g_i \left({ a^{\dag}_{i+} S_+ + a_{i+} S^{\dag}_+ + a^{\dag}_{i-} S_- + a_{i-} S^{\dag}_- }\right),
\end{eqnarray}
where $a^{\dag}_{i\pm}$ ($a_{i\pm}$) is the photon creation (annihilation) operator in the corresponding cavity mode, $S_+ = \ket{c}\bra{a}$, and $S_- = \ket{c}\bra{b}$. 

The couplings given in (3) allow to group the states of the full system composed of the atom plus the cavity modes into manifolds such that each manifold is decoupled from the rest. 
We assume the ability to prepare the intracavity fields in a Fock state \cite{Wal} and take 
$\ket{\psi(t=0)}=a^{\dag}_{1+} a^{\dag}_{1-} \ket{\Omega} (\equiv \ket{I})$ as the initial state of the system with 
$\ket{\Omega} \equiv \ket{c}\otimes \ket{\Omega}_1 \otimes \ket{\Omega}_2 $, being $\ket{\Omega}_i$ the two mode vacuum state of cavity $i$.
In this case, the coherent evolution of the system is constrained to remain in the space spanned by the five states of the manifold displayed in Fig.~2(a). Let us consider now the alternative basis of this manifold given by: 
\begin{eqnarray}
\ket{I}&\equiv&a^{\dag}_{1+} a^{\dag}_{1-} \ket{\Omega}, \\
\sqrt{2} \ket{B}& \equiv & \left( S^{\dag}_+ a^{\dag}_{1-} + S^{\dag}_- a^{\dag}_{1+} \right) \ket{\Omega}, \\
\sqrt{2} \ket{D}& \equiv & \left( S^{\dag}_+ a^{\dag}_{1-} - S^{\dag}_- a^{\dag}_{1+} \right) \ket{\Omega}, \\
\sqrt{2} \ket{E^+}& \equiv & \left( a^{\dag}_{2+}a^{\dag}_{1-}  + a^{\dag}_{2-}a^{\dag}_{1+}  \right) \ket{\Omega}, \\
\sqrt{2} \ket{E^-}& \equiv & \left( a^{\dag}_{2+}a^{\dag}_{1-}  - a^{\dag}_{2-}a^{\dag}_{1+}  \right) \ket{\Omega}. 
\end{eqnarray}
$\ket{B}$ and $\ket{D}$ are the so-called bright and dark states \cite{Ari} resulting from the combination of the excited atomic states and the modes of cavity 1. $\ket{E^{\pm}}$ correspond to two Bell states for the photons while the atomic state factorizes. In the interaction picture, it is straightforward to check that, under the two photon resonance condition, $\Delta_+ = \Delta_- (\equiv \Delta)$, the latter basis states satisfy
\begin{eqnarray}
\bra{D}H\ket{I}&=&\bra{D}H\ket{E^+}=\bra{B}H\ket{E^-}=0  \nonumber \\
\bra{B}H\ket{I}&=&\sqrt{2}g_1 e^{-i\Delta t} \nonumber \\
\bra{B}H\ket{E^+}&=&\bra{D}H\ket{E^-}=g_2 e^{-i\Delta t} \nonumber
\end{eqnarray}
The resulting coupling chain is schematically illustrated in Fig.~2(b) and suggests the two proposals of this paper.  

\begin{figure}[t]
	\centering
		\includegraphics[width=\columnwidth,clip=true]{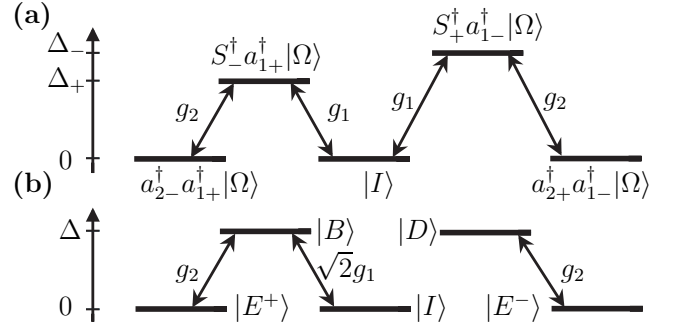}
		\caption{(a) Manifold of states coupled to $\ket{I}=a^{\dag}_{1+} a^{\dag}_{1-} \ket{\Omega}$ and the corresponding relative energies and coupling strengths (in the interaction picture). (b) The same manifold in terms of the basis states given in (4)-(8) under the two-photon resonance condition.}
	\label{fig:f2}
\end{figure}

\noindent
\textit{Proposal 1. The two ROs scheme.} Interaction starts in cavity 1 with two different pathways for the atomic excitation, from $\ket{I}$ to $S^{\dag}_{+} a^{\dag}_{1-} \ket{\Omega}$ and to $S^{\dag}_{-} a^{\dag}_{1+} \ket{\Omega}$ (see Fig.~2(a)). However, under the two-photon resonance condition, one indeed deals with an effective two-level system where ROs occur between states $\ket{I}$ and $\ket{B}$ (Fig.~2(b)). In the interaction picture, the solution of the Schr\"odinger equation for $g_2=0$ yields:
\begin{eqnarray}
\ket{\psi(t)}&=&{e^{-i\Delta t/2}} \Big{[} \frac{-i2\sqrt{2}g_{1}}{\Omega _{1}} \sin(\Omega _{1}t/2)\Big{]}\ket{B}\nonumber\\
& & \hspace{-1.3cm}+ {e^{i\Delta t/2}} \Big{[}\cos(\Omega _{1}t/2) -i \frac{\Delta}{\Omega _{1}}\sin(\Omega _{1}t/2)\Big{]}\ket{I}
\nonumber \label{wf}
\end{eqnarray}
being $\Omega_{1}=\sqrt{8g_{1}^{2}+\Delta ^{2}}$ the so-called generalized quantum Rabi frequency.
Hence, under the single-photon resonance condition, $\Delta =0$, there are complete population oscillations between these two states.

With this dynamics in mind, the steps to deterministically generate a polarization entangled photon pair are the following: (i)~preparation of the system into the initial state $\ket{I}$; (ii)~the three-level atom interacts resonantly with the two circular polarizations modes of cavity 1 for an interaction strength, $\Omega_1(t)=2\sqrt{2}g_1(t)$ \cite{tdep}, and time, $t_1$, yielding half-of-a-resonant RO with the bright state, i.e., $\int_0^{t_1 }\Omega_1(t)dt = \pi$. The state of the system after this step will be $\ket{\psi(t_1)}=-i \ket{B}$;
and, finally, (iii)~the three-level atom couples to cavity 2 for a time $t_2$ such that $\int_{t_1}^{t_1+t_2}\Omega_2(t)dt = \pi$, with $\Omega_2(t)=2g_2(t)$. If so, the vaccum modes of cavity 2 stimulate the emission of a single photon through the two paths $S^{\dag}_+ a^{\dag}_{1-} \ket{\Omega} \rightarrow a^{\dag}_{2+}a^{\dag}_{1-} \ket{\Omega}$ and $S^{\dag}_- a^{\dag}_{1+} \ket{\Omega} \rightarrow a^{\dag}_{2-}a^{\dag}_{1+} \ket{\Omega}$. The state of the system after this last step will be $\ket{\psi(t_1+t_2)} = -\ket{E^+}$.
Hence, the state of the three-level atom factorizes and, in the end, each cavity contains exactly one photon. These two photons are entangled in their polarization degree of freedom.

\smallskip
\smallskip
\noindent
\textit{Proposal 2. The STIRAP scheme.} By diagonalizing Hamiltonian (1)-(3) in the interaction picture and assuming the two-photon resonance condition, it results that one of the energy eigenstates of the system is:
\begin{equation}
\ket{\Lambda (\theta)}=\cos{\theta}\ket{I} - \sin{\theta} \ket{E^+},
\end{equation}
where the mixing angle $\theta$ is defined as $\tan \theta (t) \equiv \sqrt{2} g_1 (t) /g_2 (t)$.
Following (9), it is possible to transfer the system from $\ket{I}$ to $\ket{E^+}$ by adiabatically varying the mixing angle from $0^0$ to $90^0$
realizing a counterintuitive STIRAP process \cite{STIRAP}.
In this case, the steps to generate the polarization entangled photon pair are:  (i)~preparation of the system into the initial state $\ket{I}$; 
(ii)~the three-level atom couples first to the empty modes of cavity 2; and, before this interaction ends, (iii) the three-level atom  starts to slowly interact with the modes of cavity 1. Note that this last step means that the transverse spatial modes of the two cavities should appropriately overlap to assure the adiabaticity of the process. 

Although not as fast as the two truncated ROs proposal, the STIRAP process has two advantages: (i) it is very robust under fluctuations of the experimental parameters, e.g., interaction strengths and times, due to the fact that the system adiabatically follows an energy eigenstate; and (ii) it is almost not sensitive to atomic decay since, first, there is no need of single photon resonance, and, second, $\ket{\Lambda (\theta)}$ does never involve the intermediate state $\ket{B}$. 

To further characterize these two cQED sources of entangled photons we consider the cavity decay of the photons through the mirrors and their eventual detection. Accordingly, we will investigate next the evolution of the system in the presence of two kinds of dissipative processes. First, spontaneous atomic decay from the two optical transitions $\ket{a}$ to $\ket{c}$ and $\ket{b}$ to $\ket{c}$ at the common rate $\Gamma$. Second, cavity decay of the photons through the mirrors and the irreversible process of their detection. We assuming a perfect quantum efficiency for the detectors ($\eta =1$), $\kappa_{1\pm}=\kappa_{2\pm}\,(\equiv \kappa)$ an take the same mirror transmission coefficient, $\kappa$, for all four cavity modes. 
To account for dissipation we have used the Monte Carlo Wave Function (MCWF) formalism \cite{DCM92} and averaged over many realizations of quantum trajectories. Accordingly, Hamiltonian (1)-(3) has been replaced by the following non-hermitian Hamiltonian:
\begin{equation}
H^{\prime}=H-i{\Gamma \over 2} \sum_{i=1,2} S_i^{\dag}S_i \nonumber -i {\kappa \over 2} \sum_{i=1,2}  \left(  a_{i+}^{\dag}a_{i+} + 
 a_{i-}^{\dag}a_{i-} \right).
\end{equation}

\begin{figure}[h]
	\centering 
			\includegraphics[width=0.43\textwidth]{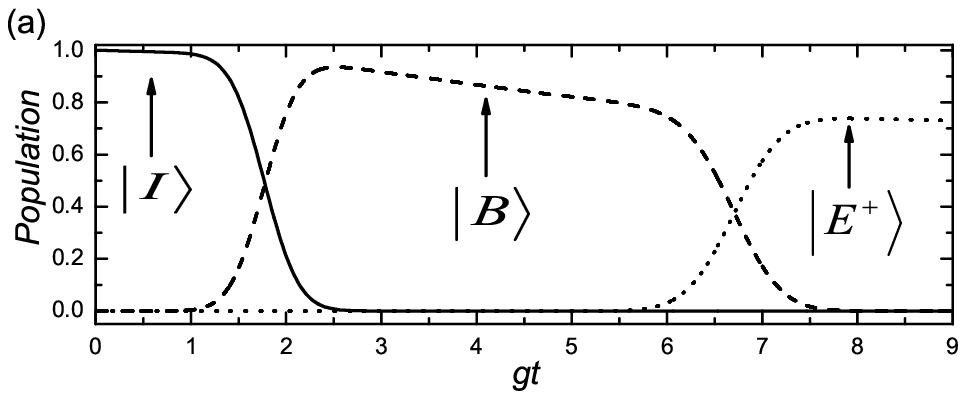}
		\includegraphics[width=0.43\textwidth]{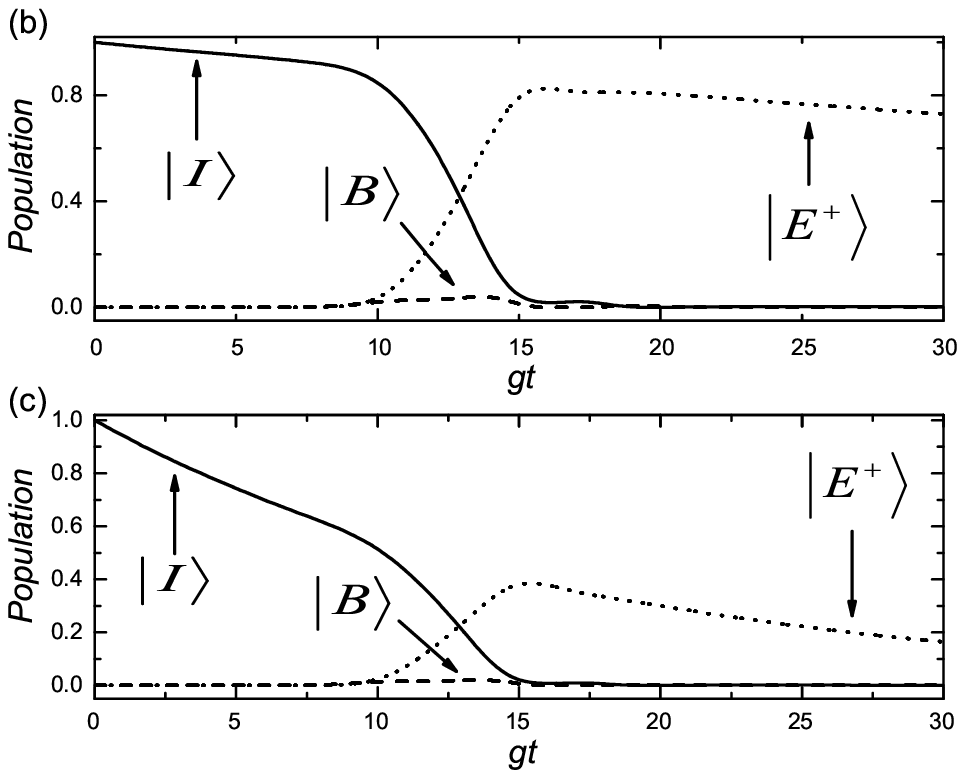}
		\caption{Evolution of the system towards the polarization entangled photon state $\ket{E^+}$ through two truncated ROs (a) and via STIRAP ((b) and (c)). Parameters are $\Gamma= 0.05g$ and $\kappa= 0.1\Gamma$ for (a) and (b); $\kappa= 0.03g$ and $\Gamma= 0.1\kappa$ for (c). In all cases $\Delta_+=\Delta_-=0$. $g$ is the vacuum Rabi frequency at the cavity center that we assume to be the same for both cavities.  Realistic parameter values have been chosen for the gaussian transverse profiles of the cavity modes. For STIRAP, appropriate overlapping between the transverse modes of the two cavities has been considered.}
	\label{fig:f3}
\end{figure}

Fig.~3 shows the evolution of the system in the presence of dissipative processes obtained by averaging over many MCWF simulations. Fig.~3(a) corresponds to the two half-of-a-resonant ROs proposal, while (b) and (c) account for the STIRAP case. In (a) and (b) the dominant dissipative process is spontaneous atomic decay, while in (c) it is cavity decay of photons through the mirrors. For (a), the fidelity of the source, defined as $F \equiv \max_t \left| \braket{E^+}{\psi(t)} \right| ^2$, is $F=0.74$, while for (b) and (c) it is $0.83$ and $0.39$, respectively. For the ROs proposal, similar results to (a) are obtained when one exchanges the values of $\kappa$ and $\Gamma$. In the STIRAP case, state $\ket{B}$ remains almost unpopulated for the whole process even with $\Delta_+=\Delta_-=0$, which makes this process quite insensitive to atomic decay (see Fig. 3(b)).  For large values of the cavity decay, the fidelity of the STIRAP scheme strongly decreases (Fig. 3(c)) since it has be to adiabatic and, therefore, significantly slow. 

To demonstrate the robustness of the STIRAP process in the presence of decoherence and experimental imperfections, in contrast to the ROs case, Fig.~4 presents contour plots of the fidelity as a function of the atomic and cavity decay rates ((a) and (b)), and of the deviation from the single and two photon resonance conditions ((c) and (d)). Figs.~4(c) and 4(d) account, e.g., for the presence of a stray magnetic field such that $(\Delta_+ - \Delta_- )/2 \neq  0 $, or an electric field yielding $(\Delta_+ + \Delta_- )/2 \neq  0 $. 
In fact, it is straightforward to check that, for a $J=0$ to $J=1$ transition, a magnetic field of one Gauss would reduce the fidelity of the cQED source by around 30\% for the ROs proposal, while in the STIRAP proposal it would be reduced by only 3\%.

\begin{figure}[t]
\vskip 0.5cm
	\centering
		\includegraphics[width=0.47\textwidth]{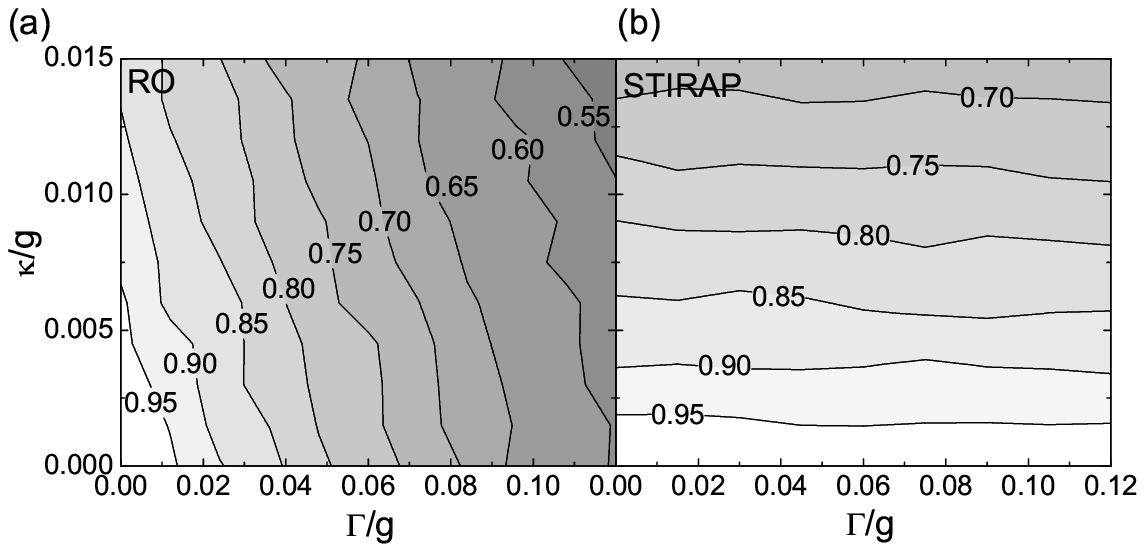}
\vskip 0.3cm
		\includegraphics[width=0.47\textwidth]{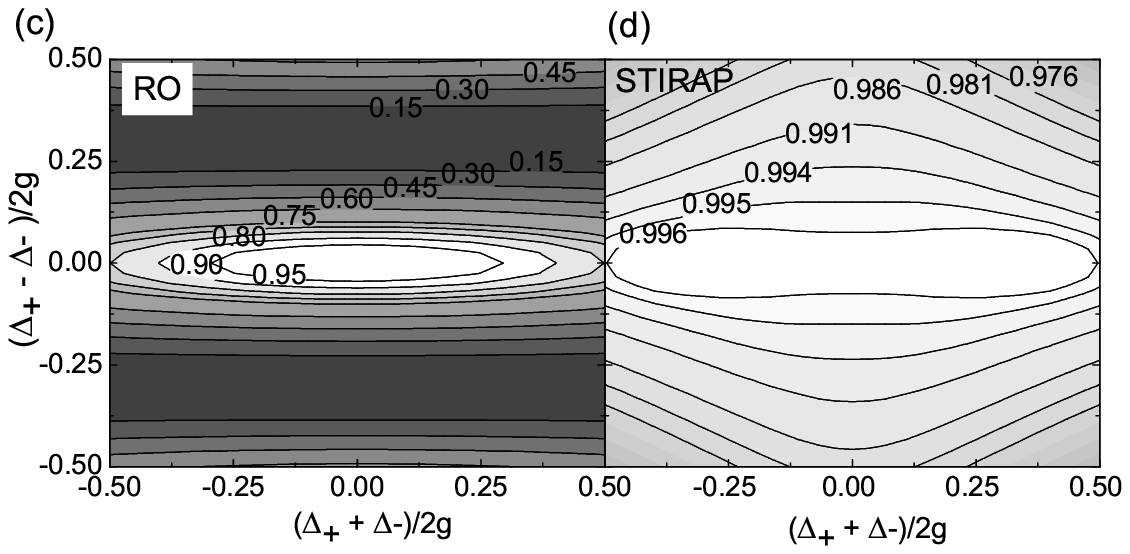}
\begin{picture}(0,0)(0,0)

\end{picture}		
		
\caption{Fidelity contour plots of the cQED entangled photon pair source for the two truncated ROs case ((a) and (c)) and via STIRAP ((b) and (d)). Parameters are: $\Delta_+=\Delta_-=0$ for (a) and (b); and $\kappa=\Gamma=0$ for (c) and (d). $(\Delta_+ + \Delta_- )/2g$ and $(\Delta_+ - \Delta_- )/2g$ measure the deviation from the single and the two-photon resonance condition, respectively. }
	\label{fig:f4}
\end{figure}

Finally, it is worth to note that by means of coincidence photodetection (see Fig.~1(b)) it is possible to discard from the statistics those processes involving spontaneous emission of photons and those where the two cavity photons have been emitted from the same cavity. For such a postselection process, Fig.~5 shows the fidelity $F$ of the cQED source and its entanglement capability, characterized by means of the $S$ parameter of the CHSH inequality \cite{CHSH} ($S=2\sqrt{2}$ for maximally entangled states and $S=\sqrt{2}$ for a non-entangled state). Clearly, for the STIRAP case, high $S$ values can be achieved even for large atomic decay rates. 

In conclusion, we have proposed two different schemes for the deterministic generation of polarization entangled photon pairs. The first proposal is based on the implementation of half-of-a-resonant RO in each cavity. Within this scheme, fidelities around $F \sim 0.4$ could be obtained for the best combination of atomic and cavity decay rates of state-of-the-art experimental implementations in the optical domain \cite{bestatomic,Sauer}. The second proposal is based on STIRAP and, although slower, it is considerably more efficient and presents interesting features such as its robustness under fluctuations of the experimental parameters or the fact that it is almost not sensitive to spontaneous atomic decay.  
For this proposal, fidelities around $F \sim 0.2$ are expected for state-of-the-art implementations.     

\begin{figure}[h]
\vskip 0.2cm
	\centering
		\includegraphics[width=0.4\textwidth]{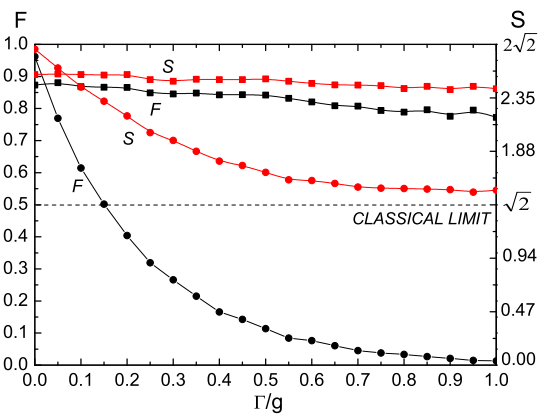}
		\caption{(Color online) Values of the fidelity $F$ and the entanglement parameter $S$ as a function of the atomic decay for the ROs method (circles) and STIRAP (squares). Parameters are, in both cases, $\Delta_+=\Delta_-=0$ and $\kappa=0.005g$. Lines are to guide the eyes.}
	\label{fig:f5}
\end{figure}

We acknowledge support from the contracts BFM2002-04369-C04-02 and FIS2005-01497MCyT (Spanish Government) and SGR2005-00358 (Catalan Government). 
KE acknowledges the support received from the European Science Foundation
(ESF) 'Quantum Degenerate Dilute Systems'.


\begin{references}
\bibitem{tdep} For the setup of Fig. 1(b), the time dependence of the couplings $g_i(t)$ is determined by the transverse profile of the cavity modes and the atomic velocity.
\end{references}
\end{document}